\documentclass[sigconf,authorversion]{acmart}
\usepackage{enumitem}
\usepackage{tabularx}
\usepackage{array}
\usepackage{subcaption}
\usepackage{multirow}
\usepackage{caption}
\usepackage[normalem]{ulem}
\usepackage{float}
\usepackage{colortbl} 
\usepackage[table]{xcolor} 
\AtBeginDocument{%
  }

\copyrightyear{2027}
\acmYear{2027}
\setcopyright{cc}
\setcctype{by}
\acmConference[SIGCSE TS 2027]{Proceedings of the 58th ACM Technical Symposium on Computer Science Education V.1}{February 17--20, 2027}{Sacramento, CA, USA}
\acmBooktitle{Proceedings of the 58th ACM Technical Symposium on Computer Science Education V.1 (SIGCSE TS 2027), February 17--20, 2027, Sacramento, CA, USA}
\acmDOI{10.1145/3857222.3860166}
\acmISBN{979-8-4007-3075-7/2027/02}

\newcommand{\todo}[1]{#1}
\newcommand{\gray}[1]{\textcolor{gray}{#1}}

\begin{document}

\title[Scaffolding GenAI Use in Ethical Essay Writing]{Scaffolding GenAI Use in Ethical Essay Writing:\\{} A Three-Year Experience Report in Introductory Computing}

\author{Asma Shakil}
\orcid{0000-0003-4760-3792}
\affiliation{%
 \institution{University of Auckland}
 \city{Auckland}
 \country{New Zealand}}
 \email{asma.shakil@auckland.ac.nz}

\author{Paul Denny}
\orcid{0000-0002-5150-9806}
\affiliation{
  \institution{University of Auckland}
  \city{Auckland}
  \country{New Zealand}
}
\email{paul@cs.auckland.ac.nz}

\author{Jun Seo}
\orcid{0000-0001-5899-2282}
\affiliation{%
 \institution{University of Auckland}
 \city{Auckland}
 \country{New Zealand}}
 \email{jun.seo@auckland.ac.nz}

 \author{Gauravkumar Hirani}
\orcid{0000-0002-5404-9534}
\affiliation{%
 \institution{Karnavati University}
 \city{Gandhinagar}
 \country{India}}
 \email{gaurav.h@karnavatiuniversity.edu.in}

 \author{Sean Longyu Ma}
\orcid{0000-0002-0854-039X}
\affiliation{%
 \institution{University of Auckland}
 \city{Auckland}
 \country{New Zealand}}
 \email{sean.ma@auckland.ac.nz}

\author{Claus Brabrand}
\orcid{0000-0001-5093-5806}
\affiliation{
  \institution{IT University of Copenhagen}
  \city{Copenhagen}
  \country{Denmark}
}
\email{brabrand@itu.dk}

\renewcommand{\shortauthors}{Asma Shakil et al.}

\begin{abstract}
As society, ethics, and professional practice become increasingly integrated across the computer science curriculum, technical courses need practical ways to support ethical reasoning. Ethical essay assignments offer one approach, but little is known about how students use GenAI when writing such essays. This experience report examines GenAI use across three years of ethical essay assignments in a large first-year computer systems course. Across six post-essay surveys from 2024 to 2026, we analyzed 1,894 responses, linked survey data with essay grades in 2025 and 2026, and introduced a scaffolded responsible-use workflow in 2026.  Under voluntary-use policies, AI adoption remained steady at around two-thirds of students,  rising above 90\% when use was scaffolded.  Students reported mainly using AI for help with grammar and style, while tool use diversified beyond ChatGPT in 2026.   Grade-linked analyses showed that AI use did not independently predict essay quality, while marker evaluations indicated high levels of critical engagement when use was scaffolded. Overall, GenAI's pedagogical value appears to depend on how assessment design guides students to evaluate and critically engage with AI-generated suggestions.
\end{abstract}


\maketitle

\section{Introduction and Background}

Generative AI (GenAI) tools have rapidly become part of the higher education student experience since the public release of ChatGPT in late 2022~\cite{prather2023robots}. In computing education, students use these tools not only for programming, but also for brainstorming, writing support, and language refinement, with studies reporting both enthusiasm and skepticism~\cite{keuning2024students, choudhuri2025insights, prather2024beyond}.

This shift coincides with the growing importance of society, ethics, and professional practice in computing education curricula. The CS2023 guidelines highlight the rising prominence of societal and ethical concerns and recommend integrating the Society, Ethics, and the Profession knowledge area across the curriculum from Year~1, including within technical courses~\cite{kumar2024cs2023}. Recent work echoes this interest while noting persistent barriers such as time constraints, limited expertise, and competing priorities~\cite{smith2023incorporating, towell2004further, pant2024teaching}. Ethical essays are a common assignment form because they require students to apply normative theories, identify stakeholders, and justify positions on contentious issues~\cite{moskal2002grading}, and GenAI heightens their relevance by introducing concerns such as algorithmic bias, misinformation, and autonomous decision-making~\cite{aler2024teach}. Short ethical essays in an introductory computing systems course therefore offer a useful setting for studying GenAI-supported writing.

However, much of the existing empirical evidence on student GenAI use comes from cross-sectional studies of single cohorts at a single point in time. While these \emph{snapshot studies} provide valuable early insights, they cannot show how student practices change as tools, policies, and pedagogies mature. Several foundational studies call for longitudinal research on changes in student behaviour over time~\cite{chan2023students, barrett2023not, faraon2025international}, a gap also noted in systematic reviews of GenAI in higher education~\cite{dos2025systematic}. \todo{Our study is not a longitudinal panel; however, it follows three successive cohorts of the same course (a repeated cross-sectional design), showing how practices changed at the course level rather than for individual students.}

Equally important is how students engage with AI output. Yang et al.~\cite{yang2025modifying} show that actively modifying AI-generated text yields higher-quality writing, whereas passively accepting suggestions does not. But this causal evidence comes from controlled settings with crowd-sourced participants, leaving open whether the same holds in authentic coursework, where students write under time constraints, on assigned topics, with varied tools.

This experience report addresses these gaps through a three-year study (2024--2026) in a large first-year introductory computing course. Across six essays, we analyse: 
\[
  N = \underbrace{448}_{2024} + \underbrace{694}_{2025} + \underbrace{752}_{2026} = \textbf{1{,}894} \text{ student responses}
\]
examining how students use GenAI for ethical essay writing, a format widely used in CS courses~\cite{moskal2002grading, brown2024teaching}. Ethical essays are a useful context because GenAI can support writing form, but also invite uncritical content and unsupported arguments. We therefore examine not only \textit{whether} students use AI, but \textit{how} they engage with it---whether they modify, reject, or passively accept suggestions. \todo{Scaffolding and AI disclosure are not new in themselves; our contribution is to track GenAI use in computing-ethics writing across three cohorts of an authentic course, link it to grades, and evaluate a draft-first scaffold through self-reports and marker coding.} We report (1) adoption patterns under voluntary (2024--2025) and scaffolded (2026) policies, (2) a reusable responsible-use scaffold, and (3) practical lessons for educators.

\subsection{Motivation and Study Arc}

The motivation for this study arose from an unexpected observation in our introductory \textit{computing systems} course. In Semester 1, 2024, we expressly allowed students ($\approx 570$) to use GenAI tools for two ethics essays, with mandatory disclosure. Contrary to expectations, many students reported not using AI at all. This gap between permission and uptake prompted our central motivating question:
\begin{quote}
    \textbf{\underline{Motivatin}g Q\underline{uestion}:} \textit{How do students actually engage with GenAI when given the freedom to use it, and how does this engagement change over time?}
\end{quote}
Our investigation unfolded in three phases (cf. Table~\ref{tab:study_overview}). 
\begin{itemize}[topsep=0pt,leftmargin=*]
\item \textbf{Phase 1 (2024): Exploration}: explored voluntary adoption, including tool choices, and purposes of use. 
\item \textbf{Phase 2 (2025): Measurement}: retained the same survey instrument but linked responses to tutor-assessed essay grades. 
\item \textbf{Phase 3 (2026): Scaffolded Redesign}: introduced a scaffolded responsible-use workflow and marker-facing engagement codes. 
\end{itemize}
This progression, from exploration to measurement to pedagogical redesign, follows a design-based research arc~\cite{barab2016design} in which each phase was informed by the findings of the previous one.

\section{Evaluation approach}
\label{sec:evaluation}
We evaluated the way that GenAI use was adopted by students for essay writing across three years of course offerings.  We were interested in understanding what changed over time, whether AI use explained essay performance, and what scaffolded engagement with AI looked like in student submissions.

We organize our evaluation around three guiding questions:

\begin{itemize}[topsep=0pt,leftmargin=*]
\item \textbf{Q1 (Evolution):} How have GenAI adoption patterns, tool preferences, purposes of use, and selected student perceptions evolved across three years of the same course?
\item \textbf{Q2 (Essay Quality):} Does GenAI use predict essay quality once prior writing ability is controlled, and do adoption trajectories predict score changes between essays?
\item \textbf{Q3 (Critical Engagement):} How do students engage with AI suggestions under scaffolded conditions, and what evidence of critical engagement is visible in their submissions?

\end{itemize}

\subsection{Context and Participants}

The study was conducted in \textit{Introduction to Computer Systems}, a mandatory first-year course in a CS degree covering standard systems topics such as binary representation, computer architecture, programming languages, operating systems, and networking. The course runs twice annually, enrolling approximately 500--600 students in the first semester and 200--300 in the second.

Participants were first-year undergraduate students enrolled in the course across three consecutive offerings. Table~\ref{tab:study_overview} summarises participation across the three phases. 

\begin{table}[t]
\centering
\begin{small}
\caption{Study overview across three phases.}
\label{tab:study_overview}
\setlength{\tabcolsep}{4pt}
\setlength{\aboverulesep}{0pt}
\setlength{\belowrulesep}{0pt}
\begin{tabular}{llll}
\toprule
\rowcolor[gray]{0.9}
\textbf{Attribute}
& \textbf{Phase 1 (2024)} 
& \textbf{Phase 2 (2025)} 
& \textbf{Phase 3 (2026)} \\
\midrule
Study: & \newline Exploration & \newline Measurement & \newline Redesign \\
\rowcolor[gray]{0.95}
\#Enrolled: & 570 & 409 & 425 \\
\#Responses: & 448 \gray{(236+212)} & 694 \gray{(360+334)} & 752 \gray{(388+364)} \\
\rowcolor[gray]{0.95}
Response Rate: & 85\% & 85\% & 90\% \\
Grade link: & No \gray{($0$)}& Yes \gray{($316$)} & Yes \gray{($342$)} \\
\rowcolor[gray]{0.95}
\#Markers: & 6 & 10 & 9 \\
AI Policy: & Optional & Optional & Scaffolded \\
\rowcolor[gray]{0.95}
Essay 1 Topic: & Digital addiction & Digital addiction & AI hiring ethics \\
Essay 2 Topic: & Job displacement & Job displacement & Soc.Media/health\\
\bottomrule
\end{tabular}
\end{small}
\end{table}

\subsection{Tasks and Assessment}

Each semester, the course includes two short essay assignments (400--500 words) on ethical issues in computing (Table~\ref{tab:study_overview}). They are the only assessment for the course's ethics learning outcome, \emph{``Discuss ethical issues arising from the use of computers in society''}, and are worth 8\% of the final grade. The essays run alongside the technical content as a separate strand rather than being tied to particular lectures. Because most students are school leavers, the essays are also their first academic writing in the CS degree. Students are asked to interpret a given prompt, construct a reasoned argument, identify stakeholders, and support their positions with credible references.

\todo{Teaching assistants marked the essays (Table~\ref{tab:study_overview}) following written marking instructions and a calibration discussion held before marking began. The 2024--2025 rubric awarded 7 points, one each for clarity, organisation, evidence, ethical reasoning, referencing, AI disclosure, and length. The 2026 rubric awarded 16 points across three levels, adding reflection (4) and weighting references (4) more heavily; the draft and disclosure form were required but unmarked. A strong essay was one that explained the topic in the student's own words, analysed stakeholders, and cited evidence; the 2026 rubric also required a clear position, two or more stakeholders, and a counterargument. Full rubrics are available at~\cite{shakil2026workflow}.}

\subsection{The 2026 Scaffolded Redesign}
The 2026 offering introduced a scaffolded workflow for both essay assignments, informed by findings from Phases 1 and 2. Each element of the workflow targeted a specific risk identified in the earlier phases (Table \ref{tab:redesign}). 
The workflow asked students to: (1)~write a complete draft independently, with a checklist at its top; (2)~ask an AI tool of their choice, using their own prompt, to refine (not rewrite) the draft; (3)~accept suggestions that improved clarity and reject any that changed their argument; and (4)~make any final edits, submit, and complete the AI disclosure form. Non-users still submitted a draft. The scaffolded workflow is publicly available~\cite{shakil2026workflow}.

\begin{table}
\centering
\caption{Workflow design rationale and scaffold details.}
\label{tab:redesign}
\renewcommand{\arraystretch}{1.2}
\definecolor{rowshade}{gray}{0.92}
\definecolor{headshade}{gray}{0.9}
\begin{footnotesize}
\begin{tabularx}{\columnwidth}{>{\raggedright\arraybackslash}p{0.36\columnwidth} X}
\toprule
\rowcolor{headshade}
{\small\textbf{Risk identified}} & {\small\textbf{Scaffolded requirement}} \\
\midrule
Students may outsource content to AI &
Mandatory independent draft before any AI use, submitted with and linked to the final essay. \\
\rowcolor{rowshade}
Students may use AI before forming an ethical argument &
Draft checklist requiring students to identify at least two stakeholders, include a counterargument, and flag claims needing evidence. \\
Students may not verify AI suggestions &
At least two credible references, with every claim needing evidence cited, as AI output may be inaccurate. \\
\rowcolor{rowshade}
Students may not reflect on AI interaction &
Structured reflection (100--200 words) on AI use, differences between draft and final essay, and potential improvements. \\
Students may not disclose AI use &
Separate AI disclosure form required; a missing draft or form resulted in zero marks. \\
\bottomrule
\end{tabularx}
\end{footnotesize}
\end{table}

To support the workflow, students submitted four parts---(P1) a final essay (50\% mark), (P2) a pre-AI draft written in their own words (not marked but mandatory), (P3) a structured reflection (25\%), and (P4) an IEEE-formatted reference list (25\%)--- along with a separately submitted (P5) AI disclosure form (Table \ref{tab:redesign}). The draft, reflection, and final essay were submitted together; hence, markers could compare all three side by side. Two scaffold elements structured how students prepared and reflected on their writing:
\begin{itemize}[topsep=2pt]
\item A \emph{four-item checklist} at the top of the draft, in which students confirmed they had (i)~written the draft in their own words, (ii)~identified at least two stakeholders, (iii)~included at least one counterargument, and (iv)~flagged claims needing supporting evidence.
\item A \emph{fixed four-prompt reflection} covering (i)~their initial reading of the prompt, (ii)~whether and how they used AI, and what was most and least useful, (iii)~the main difference between their draft and final essay, and (iv)~what they would do differently next time.
\end{itemize}
The combined submission (draft, reflection, and final essay) enabled three additional categorical codes that markers recorded:
\begin{itemize}[topsep=2pt,leftmargin=*]
\item \textbf{AI tone} (\emph{Positive}, \emph{Neutral}, \emph{Negative}, or \emph{Did not use AI}) records the marker's impression of how AI use affected the submission.
\item \textbf{Critical engagement} (\emph{Yes}, \emph{No}, or \emph{Did not use AI}) captures whether the student showed evidence of critically evaluating AI output, e.g., by describing what they rejected or modified, or by leaving visible traces of independent judgement in the final essay.
\item \textbf{Genuine insight} (\emph{Yes} or \emph{No}) captures whether the draft-to-final transition shows substantive development of the student's reasoning rather than surface-level edits.
\end{itemize}
Codes were applied to both essays ($N = 383$ for Essay 1, $N = 358$ for Essay 2 after linking with surveys). The codes were covered by the same written instructions and calibration discussion. 

\subsection{Data Collection and Analysis}

After each essay, students completed a post-essay survey via Microsoft Forms. Across all phases it captured AI use, tool choice, purposes of use, and three perception measures: trust in AI output, intention to reuse AI, and perceived improvement in essay quality. Responses were linked with essay grades in Phases~2 and 3, and in Phase~3 the survey was expanded to capture responsible-use behaviours such as reference verification and rejection or modification of AI suggestions.
All analyses are structured around the three evaluation questions in Section \ref{sec:evaluation}:

\textbf{For Q1,} we report descriptive statistics for adoption, tool, and purposes of use, and use proportional $Z$-tests to identify statistically significant cross-year shifts. 
All cross-year comparisons (except adoption rates) are based on post-Essay~1 responses, because it is the survey administered most consistently across all three offerings. 

\textbf{For Q2,} we use Phase~2 and Phase~3 grade-linked data. An analysis of covariance (ANCOVA) models Essay~2 rubric scores as the dependent variable, AI use in Essay~2 as the predictor, and Essay~1 score as a baseline covariate. We also examine score changes ($\Delta$Essay = Essay~2 $-$ Essay~1) by adoption trajectory: \begin{itemize}[topsep=0pt]
\item \textbf{Always:} used AI for \emph{both} Essay~1 \emph{as well as} Essay~2;
\item \textbf{Never:} did \emph{not} use AI for \emph{either} Essay~1 \emph{or} Essay~2;
\item \textbf{Adopt:} did \emph{not} use AI for Essay~1, but \emph{adopted} it for Essay~2;
\item \textbf{Abandon:} used AI for Essay~1, but \emph{abandoned} it for Essay~2.
\end{itemize}
\textbf{For Q3,} we focus on Phase~3 evidence of critical engagement under scaffolded use. We report student self-reports of rejection and modification as supporting evidence, and foreground the marker codes for AI tone, critical engagement, and genuine insight applied to Essay~1 and Essay~2 submissions.
All statistical tests are two-tailed and use a 95\% confidence level.

\section{Findings}

\subsection{AI Adoption, Tools, and Purposes of Use (Q1)}

Q1 examined how students' AI adoption, tool preferences, and purposes of use evolved across three course offerings. All cross-year comparisons in this section use deduplicated valid survey responses. Because the 2026 offering differed from 2024 and 2025 in that students were expected to engage with AI under a more scaffolded design, we interpret the 2026 comparisons with caution.

\begin{figure}
    \centering
    \includegraphics[width=0.95\linewidth]{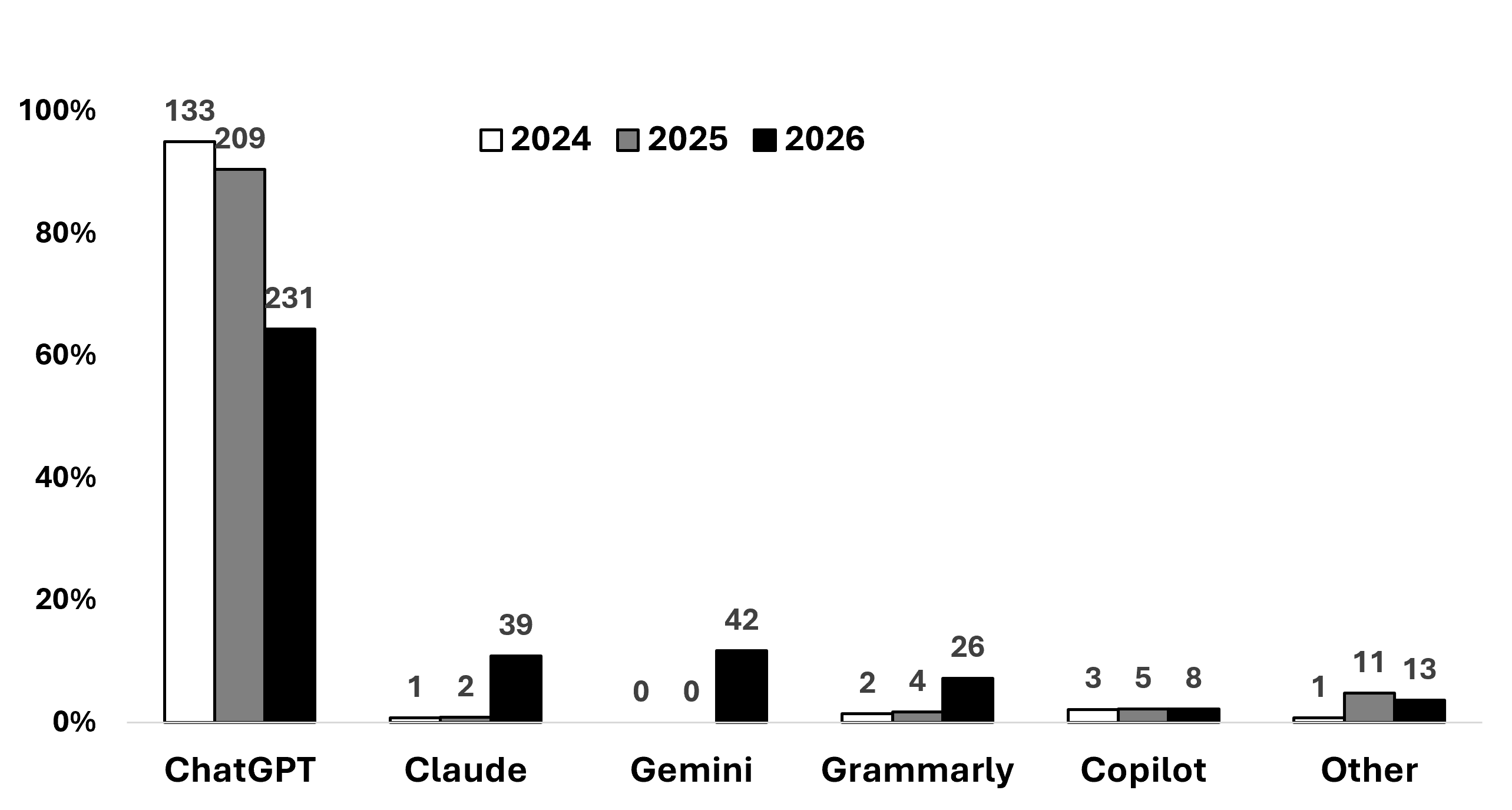}
    \caption{Primary AI tool for Essay~1, across the three phases. 
    }
    \label{fig:tools}
    \Description{Primary AI tool among AI users}
\end{figure}

\textbf{Adoption.} \todo{We define adoption as answering \emph{Yes} to ``Did you use any generative AI tool to assist in improving your essay?'', an item asked identically in all three years. It covers any AI use, whatever its purpose, and in 2026 includes the workflow's refinement step; purposes of use are reported separately below.} Under voluntary use, adoption did not change significantly: 59\% to 64\% for Essay~1 and 69\% to 68\% for Essay~2 (2024 to 2025). In 2026 it reached 93\% for both essays (both $p<.001$ vs.\ 2025). Because 2026 introduced an expected-use, structured workflow, this is not organic growth; it suggests that assessment design and policy framing strongly shape whether students use AI.
\begin{quote}
\textbf{\textsc{\underline{Result 1a}} (Policy, not time, drives adoption):} Under voluntary policy, roughly \textbf{2/3 of students} chose to use AI in both 2024 and 2025, with no significant year-on-year change. Under the 2026 expected-use policy, adoption rose sharply to \textbf{over 9/10 students} for both essays. 
\end{quote}
\textbf{Tool choice.} ChatGPT dominated in all three years but its share fell sharply in 2026 (Figure~\ref{fig:tools}). Across 2024 and 2025 it accounted for roughly 90\% of primary tool use in both essays (E1: 95\% and 91\%; E2: 92\% and 88\%; year-on-year changes non-significant), then dropped to 64\% (Essay~1) and 59\% (Essay~2) in 2026, a significant decline in both. The shift reflects diversification rather than reduced AI use: Claude and Gemini, near-absent earlier, each captured roughly one-tenth of users in 2026, and Grammarly rose to around 7\%. Copilot held steady at about 2\% throughout.

\begin{quote}
\textbf{\textsc{\underline{Result 1b}} (\textsc{ChatGPT}'s near-monopoly broken):} \textsc{ChatGPT}'s share dropped from around \textbf{90\% in 2024--2025 to around 60\% in 2026} across both essays. \textbf{Tool diversification}, not reduced AI use, drove the shift: collectively, \textsc{Claude}, \textsc{Gemini}, and \textsc{Grammarly} captured roughly 1/3 of users in 2026.
\end{quote}

\begin{figure}
    \centering
    \includegraphics[width=0.95\linewidth]{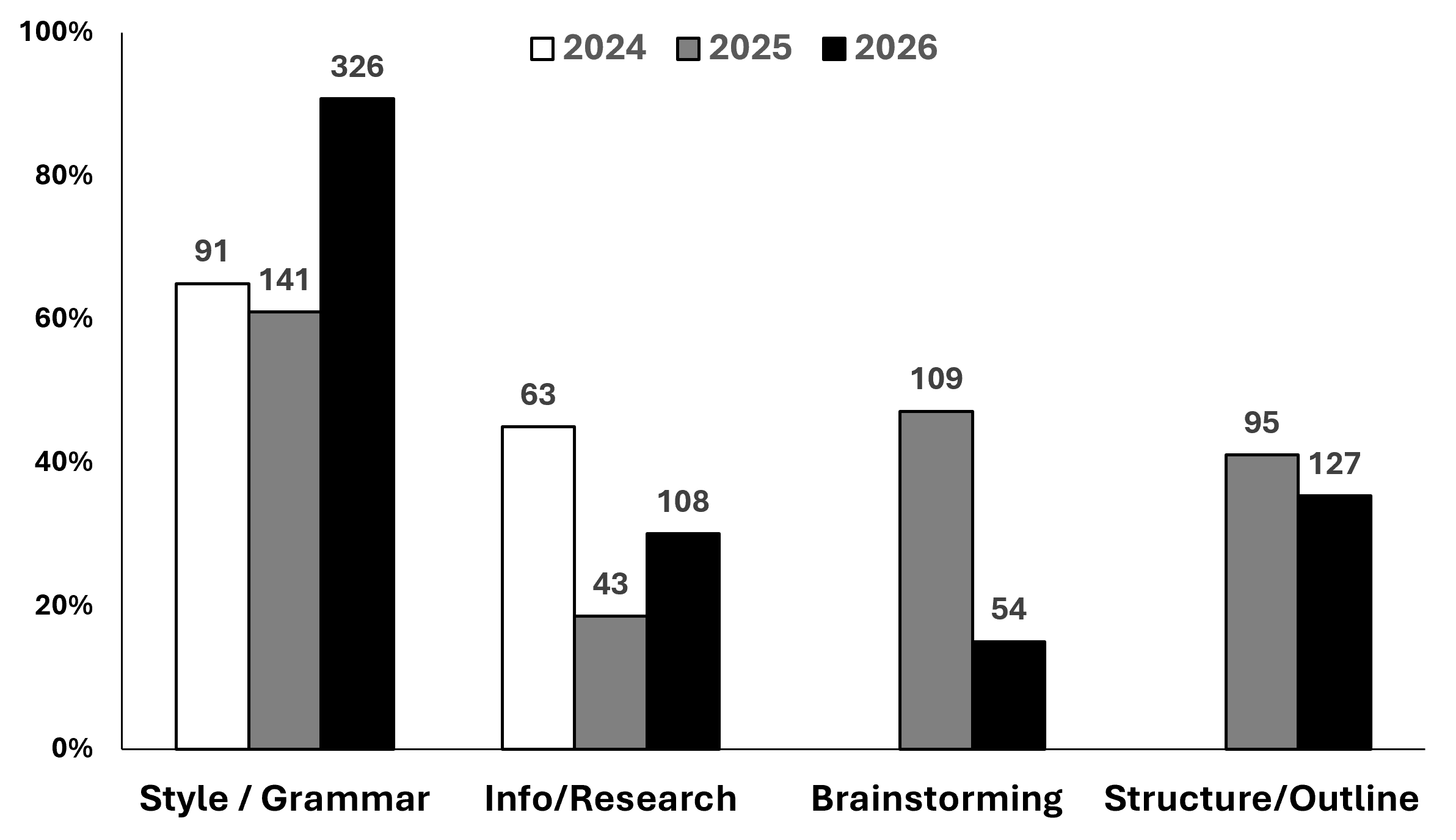}
    \caption{Self-reported task use for Essay 1}
    \label{fig:self_reported_use}
    \Description{Self-reported task use}
\end{figure}

\noindent
\textbf{Purpose of use.} Across all three phases, students used AI primarily for writing support rather than content generation (Figure~\ref{fig:self_reported_use}). 65\% of AI users reported using AI for style, grammar, and form in 2024 and 61\% in 2025, rising to 91\% in 2026---consistent with the 2026 redesign, though the changed instructional context means this is not a direct year-on-year shift. Among categories available in both 2025 and 2026, brainstorming fell sharply from 47\% to 15\%, while structure and outlining held steady (41\% to 35\%). These patterns suggest students increasingly used AI for revision and presentation rather than ideation or wholesale content generation.

\begin{figure}
    \centering
    \includegraphics[width=1.0\linewidth]{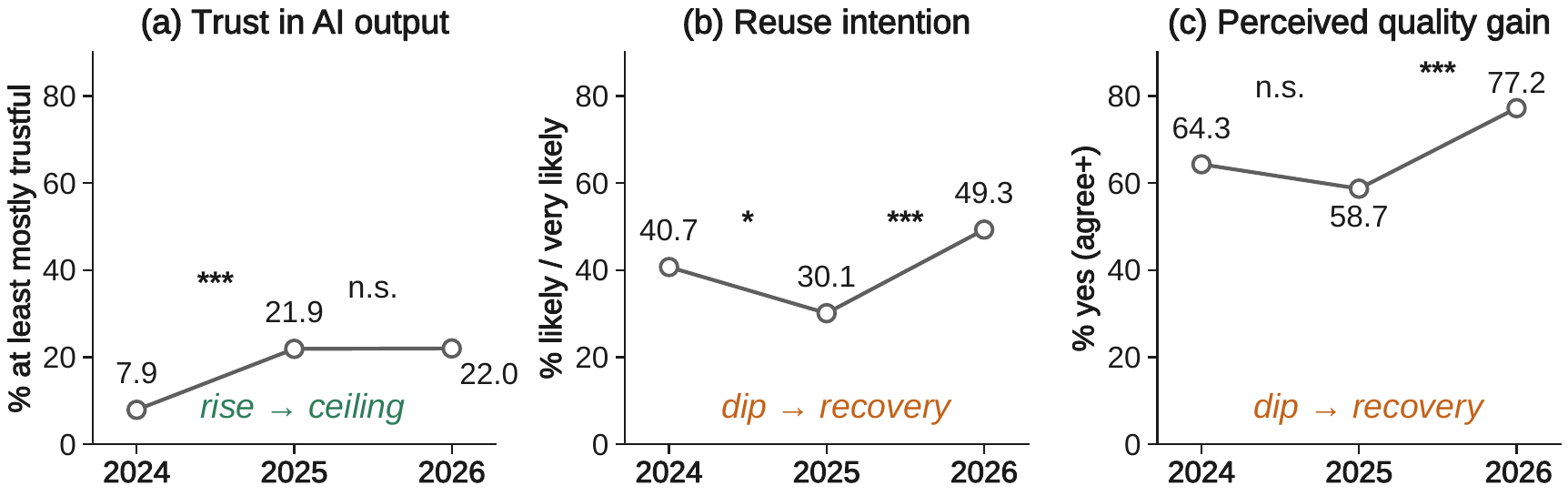}
    \caption{Changes in student perceptions among AI users across three years. Significance markers indicate year-on-year comparisons: * $p<.05$, *** $p<.001$, n.s. = not significant}
    \label{fig:trust_reuse_quality}
    \Description{Perception changes}
\end{figure}

\begin{quote}
\textbf{\textsc{\underline{Result 1c}} (Students use AI primarily for form):} Across all three years, students used AI primarily for \textbf{grammar and form} (\textsc{Essay~1}: 65\%~$\rightarrow$~61\%~$\rightarrow$~91\%). \textbf{Brainstorming fell} from 47\% in 2025 to 15\% in 2026, while structure and outlining held relatively steady.
\end{quote}
\textbf{Student perceptions.} Student perceptions did not become uniformly more positive over time (Figure~\ref{fig:trust_reuse_quality}). Trust in AI output rose from 7.9\% of AI users in 2024 to 22\% in 2025 ($p <.001$), then held at 22\% in 2026, suggesting that familiarity did not lead to uncritical acceptance. Reuse intention and perceived quality instead showed a dip-and-recovery pattern: intended reuse fell from 41\% (2024) to 30\% (2025) before rising to 49\% (2026), while perceived quality moved from 64\% to 59\% to 77\%. This may reflect early enthusiasm giving way to more measured views in 2025, then renewed confidence under the scaffolded 2026 workflow.

Taken together, Q1 shows that AI use did not simply increase with time. Under voluntary-use policies, adoption stabilised at roughly two-thirds of students; under expected and scaffolded use, adoption rose above 90\%. At the same time, the tool ecology diversified beyond ChatGPT, and students' reported purposes remained concentrated on writing support rather than content generation.

\subsection{AI Use and Essay Quality (Q2)}

Q2 examined whether AI use predicted essay quality after controlling for prior writing performance, and whether adoption trajectories between Essay~1 and Essay~2 explained changes in essay scores. Phase~2 (unscaffolded) and Phase~3 (scaffolded) yielded consistent results: AI use did not independently predict essay quality once prior performance was held constant, and adoption trajectories did not significantly explain score changes.

\textbf{ANCOVA.} We modelled Essay~2 score as a function of Essay~1 score (covariate) and AI use in Essay~2 (focal predictor) in each phase (Table~\ref{tab:phase23-ancova}). Prior writing ability was a strong predictor in both phases ($\beta=0.245$ and $0.290$, both $p<.0001$), but AI use had no independent effect (Phase~2: $\beta=0.051$, $p=.60$; Phase~3: $\beta=0.350$, $p=.30$, 95\% CI $[-0.32, +1.02]$). In Phase~3, the raw comparison agreed: AI users averaged 14.83 on the 16-point rubric versus 14.50 for non-users, non-significant under both Welch $t$-test ($p=.36$) and Mann--Whitney ($p=.17$). Neither phase produced an AI-use effect once prior writing ability was controlled.

\begin{quote}
\textbf{\textsc{\underline{Result 2a}} (No independent AI effect on essay quality):} After controlling for Essay~1 score, AI use in Essay~2 had \textbf{no significant effect} on Essay~2 quality in either phase. Prior writing ability dominated in both phases.
\end{quote}

\begin{table}[t]
\centering
\caption{ANCOVA predicting Essay~2 score from Essay~1 score and AI
use in Essay~2. Phase~2 used a 7-point rubric; Phase~3 used a 16-point
rubric.}
\label{tab:phase23-ancova}
\setlength{\aboverulesep}{0pt}
\setlength{\belowrulesep}{0pt}
\begin{tabular}{llrrrr}
\toprule
\rowcolor[gray]{0.90}
\textbf{Phase} & \textbf{Predictor} & \textbf{Coef.} & \textbf{SE}
& \textbf{$t$} & \textbf{$p$-value} \\
\midrule
Phase~2
& Essay~1 score & 0.245 & 0.051 & 4.84 & $p<0.0001$ \\
\gray{($N$=333)} & AI used in Essay~2 & 0.051 & 0.097 & 0.53 & p = 0.60 \\
\midrule
\rowcolor[gray]{0.95}
Phase~3 
& Essay~1 score & 0.290 & 0.044 & 6.58 & $p<0.0001$ \\
\rowcolor[gray]{0.95}
\gray{($N$=343)} & AI used in Essay~2 & 0.350 & 0.339 & 1.03 & p = 0.30 \\
\bottomrule
\end{tabular}
\end{table}

\begin{figure}
    \centering
    \includegraphics[width=0.9\linewidth]{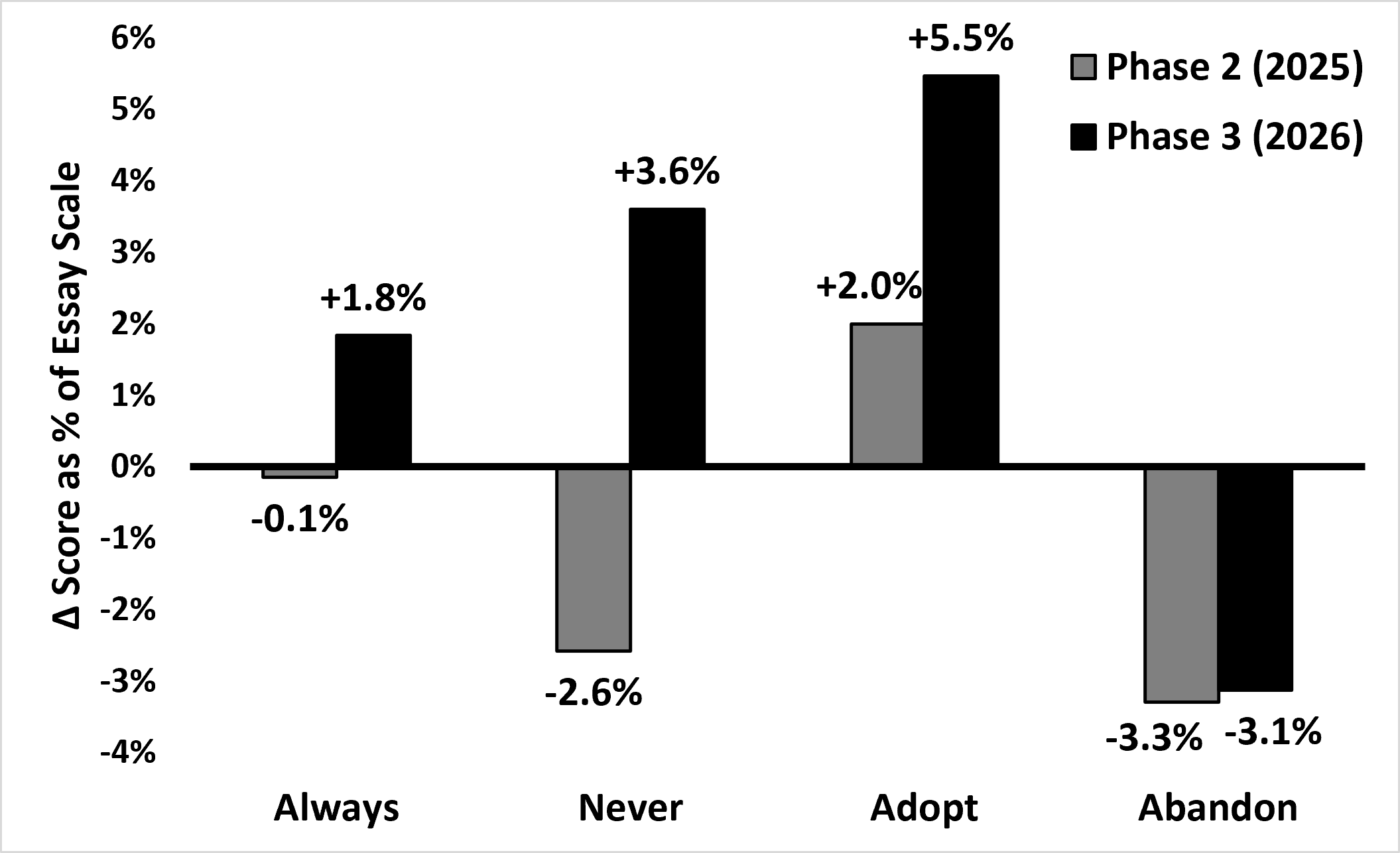}
    \caption{Mean change in essay score by AI-use trajectory for Phase 2 (2025) and Phase 3 (2026). Bars show the within-student change in essay score, calculated as $\Delta$Essay score = Essay~2 - Essay~1, expressed in \% points of the essay scale.}
    \label{fig:did_by_traj}
    \Description{Score changes by AI adoption trajectory}
\end{figure}

\noindent
\textbf{Adoption trajectories.} We grouped students into four AI-use trajectories across the two essays: Always (yes $\rightarrow$ yes), Never (no $\rightarrow$ no), Adopt (no $\rightarrow$ yes), and Abandon (yes $\rightarrow$ no). Score changes across the four groups are shown in Figure~\ref{fig:did_by_traj}. Phase~2 effects were uniformly small ($-0.23$ to $+0.14$ on the 7-point rubric), and the overall effect was not significant ($F=1.25$, $p=.29$). Phase~3 showed directionally larger effects---Adopters $+0.88$, Abandoners $-0.50$ on the 16-point rubric---but the Adopt ($n=12$) and Abandon ($n=9$) groups were too small to distinguish from noise, and the overall trajectory effect was again non-significant under one-way ANOVA ($F=0.86$, $p=.46$) and Kruskal--Wallis ($p=.51$). Neither phase supports a strong trajectory-based effect.

\begin{quote}
\textbf{\textsc{\underline{Result 2b}} (No trajectory effect in either phase):} Score changes did not differ significantly across the four adoption trajectories in either phase. In Phase~3, the Adopt ($n=12$) and Abandon ($n=9$) groups were too small to detect anything but the largest effects.
\end{quote}

\noindent
\textbf{Taken together,} Phase~2 and Phase~3 indicate that AI use---measured as a binary variable or as an adoption trajectory---does not by itself explain essay quality. Prior writing ability remained the dominant predictor across both phases. This motivates examining \textit{how} students engaged with AI, which we turn to in Q3.

\subsection{Q3: Critical Engagement With AI}

Q3 examined how students engaged with AI suggestions under the scaffolded Phase~3 workflow. The results indicate high levels of critical engagement in both self-reported practices and markers' assessments of submissions, even though AI use did not independently predict essay scores.

\textbf{Self-reported modification and verification.}
Students reported high levels of responsible AI engagement across both essays. Rejection or modification of AI suggestions was reported by 81\% of AI users in Essay 1 and 84\% in Essay 2; reference verification by 93\% (E1) and 95\% (E2). Neither change was significant, indicating a ceiling effect: students may already have understood that AI output requires verification, or may have reported responsible practices because the assessment foregrounded disclosure.

\begin{quote}
\textbf{\textsc{\underline{Result 3a}} (Self-reports showed high modification and verification):}
Rejection/modification was reported by 81\% (E1) and 84\% (E2) of AI users, and reference verification by 93\% and 95\%. 
\end{quote}
To understand \emph{why} students rejected or modified AI output, we inductively coded the open-text explanations of why students rejected or modified AI's suggestions (E1:$N=275$; E2: $N=263$). The most common reasons were protective of the students' own work rather than mechanical: \emph{preserving their own ideas, argument, or meaning} was the largest theme (E1: 39\%, E2: 34\%), followed by \emph{maintaining personal voice and writing style} (E1: 20\%, E2: 21\%). A second cluster reflected the verification the scaffold sought to encourage: \emph{factual inaccuracy or hallucination} (E1: 13\%, E2: 10\%) and \emph{citation problems} such as \emph{fabricated references} (E1: 9\%, E2: 11\%). Students described these judgments in their own words:

\begin{itemize}
\item \emph{I modified some AI suggestions because some AI ideas were different from what I wanted to argue.}
\item \emph{[It] made my writing sound too `robotic.'}
\item \emph{My opinion is much better than AI's.}
\item \emph{It told me one of my sources was wrong. It wasn't. It just hallucinated.}
\end{itemize}
These responses suggest the high rejection rates reflect deliberate,
criterion-aware judgment rather than indiscriminate refusal.
\begin{quote}
\textbf{\textsc{\underline{Result 3b}} (Rejection was deliberate, not indiscriminate):}
Open-text reasons for rejecting or modifying AI output centered on \textbf{preserving} the student's \textbf{own ideas and voice} (the largest theme) and on catching \textbf{inaccurate or fabricated references}.
\end{quote}

\noindent
\textbf{Marker-coded engagement.} In Phase~3, markers applied three new codes to both essays: \textit{AI tone}, \textit{critical engagement}, and \textit{genuine insight}. The clearest change was in AI tone (Table~\ref{tab:marker-codes}): Positive ratings increased from 47\% in Essay~1 to 66\% in Essay~2, while Negative ratings fell from 9.1\% to 0.8\% (both $p<.0001$). Critical engagement (81\% to 77\%) and genuine insight (85\% to 81\%) were already high and did not change significantly. Overall, markers found that students engaged with AI suggestions rather than accepting them uncritically.

\begin{quote}
\textbf{\textsc{\underline{Result 3c}} (Widespread critical engagement; AI tone shifted positive):} Markers coded \textbf{77--81\%} of submissions as showing critical engagement and \textbf{81--85\%} genuine insight. \textbf{Positive} AI tone rose from \textbf{47\% to 66\%}, and Negative fell from 9\% to under 1\%.
\end{quote}

\begin{table}[t]
\centering
\caption{Phase~3 marker-coded AI engagement indicators across both
essays. $N{=}383$ for Essay~1 and $N{=}358$ for Essay~2.}
\label{tab:marker-codes}
\small
\setlength{\tabcolsep}{4pt}
\setlength{\aboverulesep}{0pt}
\setlength{\belowrulesep}{0pt}
\begin{tabular}{llrrrrr}
\toprule
\rowcolor[gray]{0.9}
& & \multicolumn{2}{c}{\textbf{Essay~1}}
& \multicolumn{2}{c}{\textbf{Essay~2}} & \\
\cmidrule(lr){3-4}\cmidrule(lr){5-6}
\rowcolor[gray]{0.9}
\textbf{Code} & \textbf{Value} & $\#$ & \% & $\#$ & \% & \textbf{$p$-value} \\
\midrule
AI Tone & Positive & 181 & 47\% & 235 & 66\% & $p<0.0001$ \\
        & Neutral & 139 & 36\% & 94 & 26\% &  \\
        & Negative & 35 & 9.1\% & 3 & 0.8\% & $p<0.0001$ \\
        & Did not use AI & 24 & 6.3\% & 23 & 6.4\% &  \\
\midrule
\rowcolor[gray]{0.95}
Critical & Yes & 309 & 81\% & 275 & 77\% & $p=0.20$ \\
\rowcolor[gray]{0.95}
Engagement & No & 41 & 11\% & 56 & 16\% &  \\
\rowcolor[gray]{0.95}
         & Did not use AI & 29 & 7.6\% & 24 & 6.7\% &  \\
\midrule
Genuine & Yes & 327 & 85\% & 289 & 81\% & $p=0.09$ \\
Insight & No & 56 & 15\% & 67 & 19\% &  \\
\bottomrule
\end{tabular}
\end{table}

\noindent
\textbf{Taken together,} the Q2 and Q3 findings clarify a distinction between \emph{engagement} quality and \emph{outcome} quality. AI use did not independently predict essay scores in either phase, yet under the Phase~3 scaffold students modified, verified, and critically evaluated AI output.

\section{Lessons Learned}

This entire study was inspired by a deceptively simple question:
\begin{center}
\textbf{\emph{Why wouldn't students simply use AI when freely permitted?}}
\end{center}
Three years of data suggest a more interesting answer than \emph{they will} or \emph{they won't}. AI use is shaped by policy and assessment design as much as by tool availability; many students engage with AI selectively rather than uncritically; and AI use alone does not explain writing quality. The clearest differences under scaffolding were behavioural, changing \textit{how} students used AI rather than their scores. The lessons below summarise what we found and what we recommend:
\begin{quote}
\textbf{\textsc{\underline{Lesson 1}} (AI Adoption):} AI adoption depends on assessment framing, not only tool availability.
\end{quote}
Under stable voluntary-use policy in 2024 and 2025, AI adoption held steady at roughly two-thirds of students. In 2026, adoption exceeded 90\%, but this coincided with an expected-use and structured-workflow policy. Adoption does not rise automatically over time; it responds sharply to how instructors frame AI use in the assessment.

\begin{quote}
\textbf{\textsc{\underline{Lesson 2}} (AI Insufficiency):} AI use alone does not improve essay quality.
\end{quote}
In both Phase~2 (unscaffolded) and Phase~3 (scaffolded), ANCOVA showed no independent effect of AI use on essay quality once prior writing ability was controlled. The strongest predictor of Essay~2 score was always Essay~1 score. The binary question \emph{Did the student use AI?} is too coarse to explain writing outcomes. \todo{With several markers, some grading inconsistency is likely, which makes small effects harder to detect; this finding therefore indicates no large AI effect rather than no effect.}

\begin{quote}
\textbf{\textsc{\underline{Lesson 3}} (Scaffolding Effects):} Scaffolding shapes practice, not scores alone.
\end{quote}
Under the Phase~3 workflow, students showed high levels of modification, verification, and critical engagement, and AI tone became more positive between essays (47\% to 66\% Positive). \todo{AI use still gave no detectable score advantage, possibly because scores were near the ceiling (about 14.8/16) and AI was used mainly for grammar and style, while most marks reward reasoning, evidence, and reflection.} This is the study's central pedagogical insight: \emph{engagement quality} and \emph{outcome quality} are distinct goals, and scaffolding may advance the former without improving the latter.

\begin{quote}
\textbf{\textsc{\underline{Lesson 4}} (AI Usage Pattern):} Students mainly use AI as a writing scaffold.
\end{quote}
Across all three years, students most often used AI for form-oriented support: grammar, style, structure, clarity, and organisation. Many students did not outsource ethical reasoning to AI; they treated it as a writing assistant that helped present their ideas more clearly.
\todo{
\begin{quote}
\textbf{\textsc{\underline{Lesson 5}} (Workflow Adoption):} Keep the core scaffold; adapt the rest.
\end{quote}
}
\todo{The core elements are the draft-before-AI step with its checklist, mandatory disclosure, and a marked reflection comparing draft and final essay. Topics, word limits, weightings, tools, and referencing style can be adapted, and the engagement codes are needed only for evaluation. For CS educators, this offers a low-cost way to assess ethics within a technical course. Full materials are available at~\cite{shakil2026workflow}.}

\section{Limitations and Scope of Claims}

This experience report has several limitations. First, as a repeated cross-sectional study, it cannot track how individual students' practices evolved. Second, it was conducted in one first-year course at one institution, so the findings may not generalise to other contexts, course levels, or assessment types.
\todo{Third, the analysis relies largely on self-reported AI use, which is prone to recall and social desirability bias. Under-reporting is less likely in our study as AI use was permitted every year and expected in 2026, disclosure carried no penalty, and Phase~3 markers' coding of non-use (6--8\%) matched self-reports (7\%). Self-reported verification is more prone to over-reporting, so we treat it as supporting evidence only.}

The Phase~3 redesign also limits causal interpretation. Both 2026 essays used the scaffolded workflow, so there was no unscaffolded control, and cross-year differences mix design effects with cohort effects. Phases~2 and 3 used different score scales, so their grade-linked findings are comparable only conceptually. The engagement codes lack reliability testing, and high Phase~3 scores suggest ceiling effects. We therefore treat the findings as formative evidence of how scaffolding shaped students' AI practices, not as causal evidence that it improves essay performance.

\section{Conclusion}

This experience report examined the use of GenAI for ethical essay writing over three years in a large first-year computer systems course. Across 1,894 survey responses, adoption held at about two-thirds under voluntary use (2024--2025) and exceeded 90\% under scaffolded, expected use (2026). Students mainly used AI for writing support, and tool use diversified beyond ChatGPT.

In both grade-linked phases, neither AI use nor adoption trajectory independently explained essay quality, supporting evidence that use/non-use comparisons are insufficient~\cite{yang2025modifying}. The clearest Phase~3 findings were instead behavioural, with students reporting high levels of modification and verification, and markers saw widespread critical engagement and genuine insight.

The value of GenAI in writing-intensive computing assignments therefore appears to lie less in permitting its use than in designing structures, such as draft-before-AI workflows, checklists, reflection, and disclosure, that help students evaluate, adapt, and take responsibility for AI suggestions.

\bibliographystyle{ACM-Reference-Format}
\balance 
\bibliography{SIGCSE-Citation}
\end{document}